\documentclass[prd,twocolumn,showpacs,superscriptaddress,nofootinbib,floatfix,showkeys,10pt]{revtex4-2}
\usepackage{bm}
\usepackage{times}
\usepackage{braket}
\usepackage{amsfonts,amssymb,stmaryrd,latexsym,amsmath}
\usepackage[usenames,dvipsnames]{color}
\usepackage{epsfig}
\usepackage{slashed}
\usepackage{hyperref}
\usepackage{subfigure}
\usepackage{graphicx}
\usepackage{slashed}
\usepackage{orcidlink}
\usepackage{multirow}
\usepackage{appendix}
\usepackage{dashrule}
\usepackage{bbm}
\usepackage[normalem]{ulem}
\usepackage{mdframed}

\allowdisplaybreaks[1]

\definecolor{nicered}{rgb}{0.7,0.1,0.1}
\definecolor{nicegreen}{rgb}{0.1,0.5,0.1}
\definecolor{emph}{rgb}{1,0,0}
\definecolor{doub}{rgb}{0.7,0.2,1.0}
\definecolor{navyblue}{RGB}{0, 110, 184}
\hypersetup{colorlinks,citecolor=nicegreen,linkcolor=nicered,urlcolor=navyblue}

 \newcommand{\clabel}[2][]{#2}
 \newcommand{\change}[1]{#1}

\begin{document}

	
	\title{Meson-nucleus bound states with neural-network quantum states} 
	\author{Wei-Lin Wu\,\orcidlink{0009-0009-3480-8810}}\email{wlwu@pku.edu.cn}
	\affiliation{School of Physics, Peking University, Beijing 100871, China}
	\author{Liang-Zhen Wen\,\orcidlink{0009-0006-8266-5840}}\email{wenlzh\_hep-th@stu.pku.edu.cn}	
    \affiliation{School of Physics, Peking University, Beijing 100871, China}
	\author{Shi-Lin Zhu\,\orcidlink{0000-0002-4055-6906}}\email{zhusl@pku.edu.cn}
	\affiliation{School of Physics and Center of High Energy Physics,
		Peking University, Beijing 100871, China}
	
	\begin{abstract}
We present the first systematic calculations of $\phi$-, $\eta_c$-, and $J/\psi$-nucleus ground states up to mass number $A{=}12$ based on the HAL QCD meson-nucleon potentials at near-physical point. The $(A{+}1)$-body Schr\"odinger equation is solved with a neural-network variational Monte Carlo framework, generalized to incorporate mesonic degrees of freedom. Benchmarking on light nuclei from $^2$H to $^{12}$C yields ground-state energies consistent with the experiment. Meson-nucleus bound states emerge at $A\ge2$ for $\phi$, $A\ge4$ for $J/\psi$, and $A\ge6$ for $\eta_c$.  The $\phi$-nucleus systems exhibit the strongest binding, with binding energies reaching tens of MeV. The $J/\psi$-nucleus and $\eta_c$-nucleus systems are weakly bound at the few-MeV and sub-MeV scale, respectively. The binding energy per nucleon deepens nearly linearly with $A$ for charmonium systems, whereas the $\phi$-nucleus system exhibits a non-monotonic behavior peaking at $^4$He---a distinctive hallmark of the short-range and strongly attractive $\phi N$ interaction. The meson compresses the nucleon distribution relative to the parent nucleus, and evolves from  a halo configuration to one embedded inside the nucleus with increasing $A$. Our results provide predictions for future experimental searches, and establish a quantitative bridge between lattice QCD meson-nucleon interactions and the emergent many-body phenomena in meson-nucleus bound states.
	\end{abstract}
	\maketitle

    \section{Introduction}
    Understanding the interactions between hadrons is a central problem in particle physics.  A particularly interesting class is the interaction between quarkonium $M$ ($c\bar c$, $s\bar s$) and nucleon $N$, which share no valence quarks. Single-meson exchange is suppressed by the Okubo-Zweig-Iizuka rule, so the leading contribution may arise from short-range multi-gluon exchange~\cite{Brodsky:1989jd}. The $MN$ interaction therefore provides a unique window into gluonic dynamics, revealing different aspects of the strong force compared with the $NN$ interaction. Moreover, the $MN$ system is closely tied to exotic hadron spectroscopy: $J/\psi p$ is the discovery channel of the hidden-charm pentaquark states $P_c$~\cite{LHCb:2015yax,LHCb:2019kea}, and the $\phi p$ channel has been proposed to host their strange analogues~\cite{Wu:2023ywu,Tian:2025bkx}.

    The meson-nucleus bound states serve as a testing ground to study the $MN$ interaction. Brodsky et al.~\cite{Brodsky:1989jd} first predicted $\eta_c$-nucleus bound states for $A \geq 3$ using the QCD van der Waals force picture~\cite{Appelquist:1978rt}. Subsequent works adopted various frameworks~\cite{Wasson:1991fb,Luke:1992tm,Kaidalov:1992hd,deTeramond:1997ny,Gao:2000az,Sibirtsev:2005ex,Huang:2005gw,Belyaev:2006vn,Tsushima:2011kh,Wu:2012wta,Yokota:2013sfa,Beane:2014sda,Cobos-Martinez:2017woo}, yielding widely divergent predictions for the binding energies. Although no meson-nucleus bound state has been observed experimentally, photoproduction cross sections of $\gamma p \to \phi p$~\cite{LEPS:2005hax,Dey:2014tfa}, $\gamma p \to J/\psi p$~\cite{GlueX:2019mkq,GlueX:2023pev,Duran:2022xag}, and $J/\psi$ off light nuclei~\cite{Pybus:2024ifi}, together with the $p$-$\phi$ correlation function~\cite{ALICE:2021cpv}, have been measured. Analyses of these data support attractive $\phi N$ and $J/\psi N$ interactions, yet the extracted two-body scattering lengths span several orders of magnitude depending on the analysis framework~\cite{Strakovsky:2020uqs,Gryniuk:2016mpk,Strakovsky:2019bev,JointPhysicsAnalysisCenter:2023qgg,Chizzali:2022pjd,Wu:2024xwy,Strakovsky:2025rsm}. The discrepancies in the predicted spectrum and two-body scattering lengths call for a systematic, first-principles study of meson-nucleus bound states, which is crucial for understanding the underlying dynamics and guiding future experimental searches.

    Recently, HAL QCD Collaboration extracted the $\phi N(^4S_{3/2})$~\cite{Lyu:2022imf} and charmonium-$N$~\cite{Lyu:2024ttm} potentials from lattice QCD at near-physical pion mass, providing first-principles input for calculations of meson-nucleus bound states. Some subsequent studies have employed these potentials~\cite{Filikhin:2024avj,Wen:2025wit,Etminan:2025yoq,Zhou:2025anp}, but they are restricted to three-body systems or rely on $\alpha$-cluster approximations, introducing uncontrolled model uncertainty. A full dynamical treatment of meson-nucleus systems across light nuclei is still missing. Such calculations are challenging, as they require solving the $(A{+}1)$-body Schr\"odinger equation with short-range interactions while accommodating possible halo meson configurations.
    
    The neural-network-based variational Monte Carlo (NN-VMC) approach is well suited to such a many-body problem. Owing to the capacity of deep neural networks to approximate high-dimensional functions, neural-network quantum states have emerged as a powerful many-body wave-function ansatz~\cite{Carleo:2016svm}. The VMC approach optimizes the network parameters by minimizing the energy expectation value, so that the trial wave function converges to the ground state. NN-VMC has been successfully applied to solve electronic structures~\cite{Pfau:2019xup,Hermann:2020xqs}, nuclear ground states~\cite{Adams:2020aax,Gnech:2021wfn,Yang:2022rlw,Gnech:2023prs}, hypernuclei~\cite{DiDonna:2025oqf,Zhang:2025okd}, and multiquark states~\cite{Wu:2025wvv}, reaching accuracies competitive with or surpassing established many-body methods, with computational costs that scale polynomially with the particle number. These features enable accurate solutions of meson-nucleus bound states without dynamical approximations commonly used to simplify many-body calculations. 
    
    In this work, we carry out the first calculations of $\phi$-, $\eta_c$-, and $J/\psi$-nucleus ground states up to $A{=}12$ based on the HAL QCD interactions using NN-VMC framework. We first benchmark the framework on light nuclei from ${}^2$H to ${}^{12}$C, then perform a systematic study of the meson-nucleus bound states, examining their binding energies, spatial structures, and spin splittings, and discussing how the results reflect the underlying meson-nucleon dynamics.

	\section{\protect\clabel[hamiltonian]{Hamiltonian}}
	The nuclear interactions in this work are described by a leading-order pionless effective field theory (LO $\slashed{\pi}$EFT)~\cite{Hammer:2019poc}, which assumes that nucleon momenta are well below the pion mass, so that
	pion exchange is absorbed into short-range contact interactions.
	\begin{equation}\label{eq:VN}
		V_N = \sum_{i<j} (v^{\mathrm{CI}}_{ij}+v^{\mathrm{EM}}_{ij})
		+ \sum_{i<j<k} V_{ijk}\,.
	\end{equation}
	The charge-independent (CI) two-nucleon potential takes the form
	\begin{equation}\label{eq:vCI}
		v^{\mathrm{CI}} = C_{01}C_1(r)\,P_0^\sigma P_1^\tau
		+ C_{10}C_0(r)\,P_1^\sigma P_0^\tau\,,
	\end{equation}
	where $P_S^\sigma$ ($P_T^\tau$) is the spin (isospin) projector onto
	total spin (isospin) $S$ ($T$), and $C_{01}$, $C_{10}$ are the
	low-energy constants (LECs) for the ${}^1S_0$ and ${}^3S_1$
	channels, respectively. $C_i(r)=\frac{1}{\pi^{3/2}R_i^3}e^{-(r/R_i)^2}$ is the Gaussian regularized function. The electromagnetic term includes only the finite-size proton--proton Coulomb repulsion~\cite{Wiringa:1994wb},
	\begin{equation}\label{eq:vEM_sup}
		v^{\mathrm{EM}}(r_{ij}) = \alpha\,\frac{F_c(r_{ij})}{r_{ij}}\,
		\frac{(1+\tau_{iz})(1+\tau_{jz})}{4},
	\end{equation}
	where $\alpha$ is the fine-structure constant, and the isospin projector restricts the interaction to proton--proton pairs. The finite-size form factor is 
	\begin{equation}\label{eq:Fc_sup}
		F_c(r) = 1 - \Bigl(1 + \tfrac{11}{16}br + \tfrac{3}{16}(br)^2 + \tfrac{1}{48}(br)^3\Bigr) e^{-br},
	\end{equation}
	with $b = 4.27$ fm$^{-1}$. The three-nucleon force is a
	Gaussian contact term,
	\begin{equation}\label{eq:3NF}
		V_{ijk} = \frac{c_E}{f_\pi^4\,\Lambda_\chi}\,
		\frac{(\hbar c)^6}{\pi^3 R_3^6}\,
		\sum_{\mathrm{cyc}} e^{-(r_{ij}^2+r_{jk}^2)/R_3^2}\,,
	\end{equation}
	All parameters are taken from the model-``o'' fit of
	Ref.~\cite{Schiavilla:2021dun}, where the LECs and cutoff parameters were fitted to reproduce the $np$ effective-range expansion and the binding energies of ${}^2$H and ${}^3$H.
	
\begin{figure}[htbp]
		\centering
		\includegraphics[width=0.9\columnwidth]{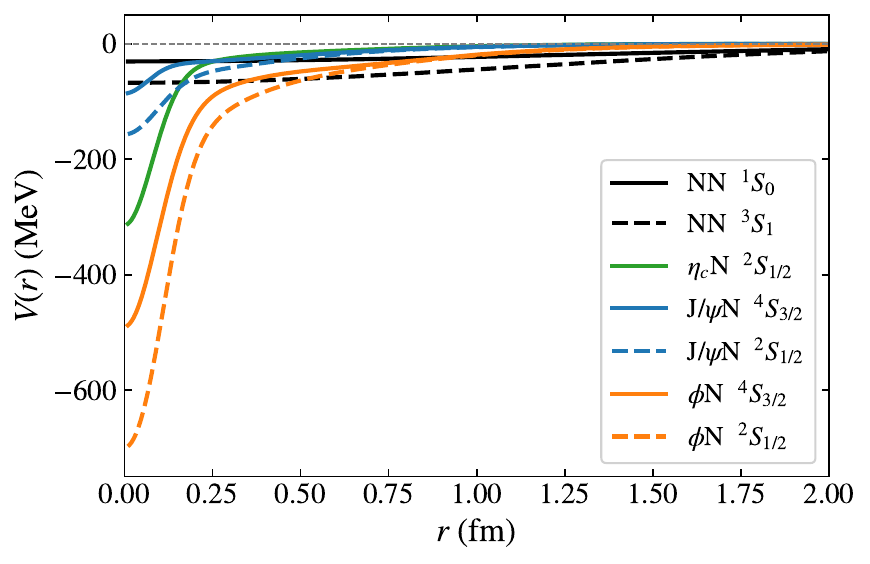}
		\caption{The $NN$ potentials in ${}^1S_0$ and ${}^3S_1$ channels of the LO $\slashed{\pi}$EFT (black), and the HAL QCD meson--nucleon potentials for $\eta_c$-$N$ (green), $J/\psi$-$N$ (blue), and $\phi$-$N$ (orange) in the ${}^4S_{3/2}$ (solid) and ${}^2S_{1/2}$ (dashed) channels.}
		\label{fig:potentials}
	\end{figure}

	The meson--nucleon interactions are described by the HAL QCD potential at $m_\pi=146$ MeV~\cite{Lyu:2022imf, Lyu:2024ttm}. The $J/\psi$-$N$ and $\eta_c$-$N$ potentials are parametrized as sums of three Gaussian functions,
	\begin{equation}\label{eq:MN_gauss}
		\begin{aligned}
			& V_{J/\psi N}^{{}^4S_{3/2}}(r)=\sum_{i=1}^3 a_{i1} e^{-\left(r/b_{i1}\right)^2}, \\
			& V_{J/\psi N}^{{}^2S_{1/2}}(r)=\sum_{i=1}^3 a_{i2} e^{-\left(r/b_{i2}\right)^2}, \\
			& V_{\eta_c N}^{{}^2S_{1/2}}(r)=\sum_{i=1}^3 a_{i3} e^{-\left(r/b_{i3}\right)^2}.
		\end{aligned}
	\end{equation}
	The $\phi$-$N$ potential in the ${}^4S_{3/2}$ channel includes two Gaussian functions and a two-pion exchange tail,
	\begin{equation}\label{eq:phiN_quartet}
		V^{{}^4S_{3/2}}_{\phi N}(r) = \sum_{i=1}^{2} \alpha_i\, e^{-(r/\beta_i)^2} + \alpha_3 m_\pi^4\, f(r, \beta_3) \left(\frac{e^{-m_\pi r}}{r}\right)^2.
	\end{equation}
	In the ${}^2S_{1/2}$ $\phi$-$N$ channel, no direct lattice QCD potential is available due to the complex coupled-channel effects. We construct the potential by assuming the meson--nucleon potential can be decomposed as
	\begin{equation}\label{eq:vMN}
		V_{MN}(r) = V^c(r) + V^{s}(r)\,\vec{S}_M\!\cdot\!\vec{S}_N\,,
	\end{equation}
	where $V^c$ and $V^{s}$ are the central and
	spin--spin components. We further assume that the spin-dependent interactions in the $J/\psi$-$N$ and $\phi$-$N$ potentials are inversely proportional to the meson mass. The ${}^2S_{1/2}$ $\phi$-$N$ potential is then obtained as~\cite{Wen:2025wit}
	\begin{equation}\label{eq:phiN_doublet}
		\begin{aligned}
			&V^{s}_{\phi N}(r) = \frac{m_{J/\psi}}{m_\phi}\, V^{s}_{J/\psi N}(r)=\frac{m_{J/\psi}}{m_\phi}\,\frac{2(V^{{}^4S_{3/2}}_{J/\psi N} - V^{{}^2S_{1/2}}_{J/\psi N})}{3},\\
			&V^{{}^2S_{1/2}}_{\phi N}(r) = V^{{}^4S_{3/2}}_{\phi N}(r) - \frac{3}{2}\, V^{s}_{\phi N}(r).
		\end{aligned}
	\end{equation}
	The full Hamiltonian reads,
	\begin{equation}\label{eq:Hamiltonian}
		H=\sum_i \frac{-\nabla_i^2}{2m_N}+\frac{-\nabla_M^2}{2m_M}+V_N+V_{MN},
	\end{equation}
	where $m_N$ and $m_M$ are the nucleon and meson masses in lattice configurations with $m_\pi=146$ MeV, as listed in Table~\ref{tab:masses}. It is worth noting that they lie close to the physical values. The two-body $NN$ and $MN$ potentials are shown in Fig.~\ref{fig:potentials}. 
	
	\begin{table}[htbp]
		\centering
		\caption{The hadron masses in lattice configurations with $m_\pi = 146\;\mathrm{MeV}$~\cite{Lyu:2022imf,Lyu:2024ttm},
			together with the corresponding experimental
			values~\cite{ParticleDataGroup:2024cfk}.
			For charmonia, Ref.~\cite{Lyu:2024ttm} uses two sets of parameters to interpolate the physical charm quark mass, and we therefore adopt the experimental masses for $J/\psi$ and
			$\eta_c$ in this work.}
		\label{tab:masses}
		\begin{tabular}{lcc}
			\hline\hline
			Hadron & This work (MeV) & Expt.\ (MeV) \\
			\hline
			$\pi$     & $146$                     & $138.0$ \\
			$N$       & $954$                   & $938.9$ \\
			$\phi$    & $1048$                         & $1019.5$ \\
			$J/\psi$  & $3096$ & $3096.9$ \\
			$\eta_c$  & $2984$ & $2984.1$ \\
			\hline\hline
		\end{tabular}
	\end{table}

	\section{\protect\clabel[nnwf]{Neural-network wave function}}
\begin{figure*}[htbp]
		\centering
		
		\includegraphics[width=0.85\textwidth]{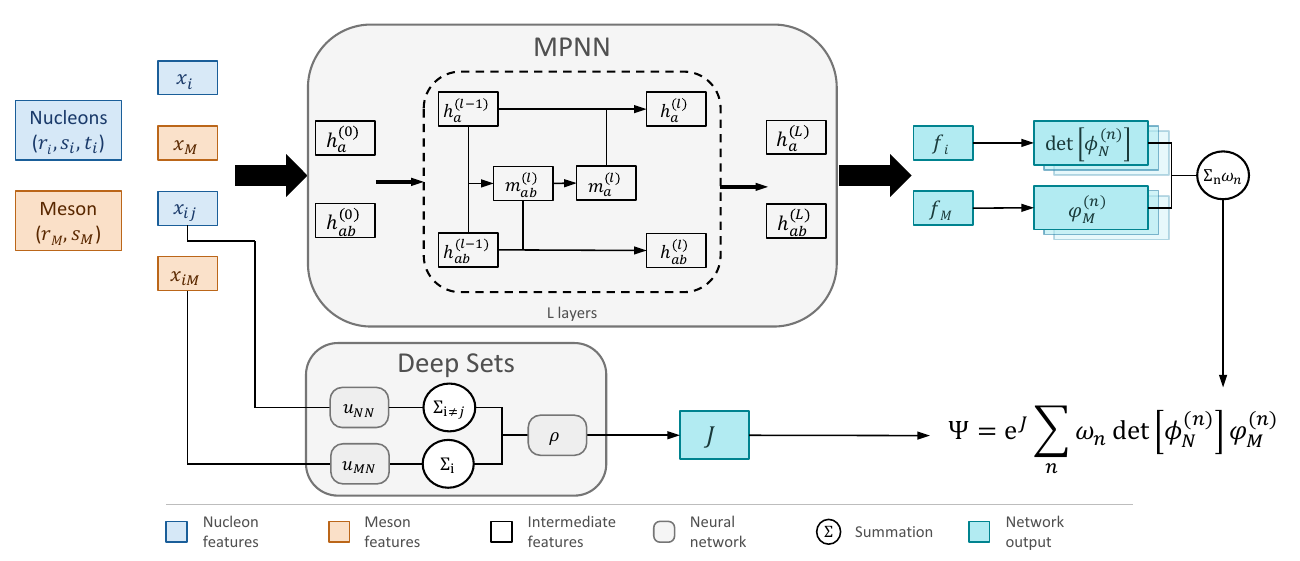}
		\caption{Architecture of the meson--nucleus neural-network wave function. An MPNN produces many-body dressed features $\boldsymbol{f}_i$ and $\boldsymbol{f}_M$, which are further transformed into nucleonic and mesonic orbitals $\phi_N$ and $\varphi_M$. The Deep Sets module gives a permutation-invariant Jastrow factor $J$. The wave function is constructed in the Slater--Jastrow form. }
		\label{fig:wf}
	\end{figure*}

	The neural-network trial wave function adopts the Slater-Jastrow form. A Deep Sets module~\cite{Zaheer:2017wmg,Wagstaff:2019limit,Adams:2020aax} constructs a permutation-invariant Jastrow correlation factor. A message-passing neural network (MPNN)~\cite{Gilmer:2017arh,Pescia:2023mcc} generates backflow-transformed nucleonic and mesonic orbitals that encode many-body correlations. Slater determinants of the nucleonic orbitals enforce antisymmetry, while the mesonic orbitals enter as additional factors. The architecture is illustrated in Fig.~\ref{fig:wf}. 

	The inputs of the neural networks are constructed as one-particle and two-particle features $\boldsymbol{x}_i = (\bar{\boldsymbol{r}}_i,\, s_i,\, t_i)$, $\boldsymbol{x}_{ij} = (\boldsymbol{r}_{ij},\, r_{ij},\, s_i, s_j, t_i, t_j)$,
	where $\bar{\boldsymbol{r}}_i$ is the nucleon coordinate in the center-of-mass (c.m.) frame, $\boldsymbol{r}_{ij} = \bar{\boldsymbol{r}}_i - \bar{\boldsymbol{r}}_j$, ${r}_{ij} = |\boldsymbol{r}_{ij}|$, and $s_i, t_i$ are the nucleon spin and isospin projections along the $z$ axis. Mesonic features are defined analogously. Working in the c.m. frame ensures translational invariance of the wave function and removes the c.m. motion.
	
	The permutation-invariant Jastrow correlation factor is written in the Deep Sets form,
	\begin{equation}\label{eq:jastrow}
		J = \rho\!\left(\sum_{i\neq j}u_{NN}(\boldsymbol{x}_{ij}),\;\;\sum_i u_{MN}(\boldsymbol{x}_{iM})\right),
	\end{equation}
	where $u_{NN}$ and $u_{MN}$ are residual networks~\cite{He:2015wrn} embedding two-particle features into a latent space $\mathbb{R}^{F}$, and $\rho: \mathbb{R}^{2F}\!\to\mathbb{R}$ decodes the pooled features into a scalar Jastrow. 
	
	The backflow transformation~\cite{Luo:2019iaq,Pfau:2019xup,Hermann:2020xqs,Yang:2022rlw} generalizes the single-particle orbitals to functions depending on the features of all particles $\phi_{i\mu}(\boldsymbol{x}_1,\dots,\boldsymbol{x}_A,\boldsymbol{x}_M)$, encoding inter-particle correlations within the determinant. Antisymmetry requires that nucleon exchange $i\leftrightarrow j$ swaps the corresponding orbital rows $\phi_{i\mu}$ and $\phi_{j\mu}$, while leaving the remaining rows unchanged, thereby flipping the sign of $\det\phi$. The construction proceeds in two stages.
	
	First, an MPNN maps the input features $\boldsymbol{x}_i$, $\boldsymbol{x}_M$, $\boldsymbol{x}_{ij}$, $\boldsymbol{x}_{iM}$ into many-body dressed single-particle features $\boldsymbol{f}_i$ and $\boldsymbol{f}_M$. Linear transformations $\boldsymbol{h}^{(0)}=W\boldsymbol{x}+\boldsymbol{b}$ project the input features into hidden features of dimension $F$, with $W$ and $\boldsymbol{b}$ being learnable parameters. The hidden features are then updated as
	\begin{equation}\label{eq:mpnn}
		\begin{aligned}
			\boldsymbol{m}_{ab}^{(l)} &= M^{(l)}\!\bigl(\boldsymbol{h}_a^{(l-1)},\boldsymbol{h}_b^{(l-1)},\boldsymbol{h}_{ab}^{(l-1)}\bigr)  ,\\
			\boldsymbol{m}_a^{(l)} &= \sum_{b\neq a}\boldsymbol{h}_b^{(l-1)} \odot \boldsymbol{m}_{ab}^{(l)} ,\\
			\boldsymbol{h}_a^{(l)} &= \boldsymbol{h}^{(l-1)}_a + U^{(l)}\bigl(\boldsymbol{h}^{(l-1)}_a,\boldsymbol{m}^{(l)}_a\bigr),\\
			\boldsymbol{h}^{(l)}_{ab} &= \boldsymbol{h}^{(l-1)}_{ab} + V^{(l)}\bigl(\boldsymbol{h}^{(l-1)}_{ab},\boldsymbol{m}^{(l)}_{ab}\bigr).
		\end{aligned}
	\end{equation}
	Here, $\boldsymbol{m}_{ab}^{(l)}$, $\boldsymbol{m}_a^{(l)} \in \mathbb{R}^F$ are the message features, $M^{(l)},U^{(l)},V^{(l)}$ are fully connected neural networks, $\odot$ denotes the Hadamard product, and the indices $a,b$ run over the $A+1$ particles. After $L$ iterations, the many-body dressed features $\boldsymbol{f}_i$ and $\boldsymbol{f}_M$ are constructed as
	\begin{equation}\label{eq:mpnn_pool}
		\boldsymbol{f}_a = \boldsymbol{h}_a^{(L)} + \sum_{b=1}^{A+1}\lambda\bigl(\boldsymbol{h}_b^{(L)}\bigr),
	\end{equation}
	with $\lambda$ being a residual network. This permutation-equivariant MPNN structure ensures that the nucleon exchange $i\leftrightarrow j$ amounts to permuting of output features $\boldsymbol{f}_i\leftrightarrow \boldsymbol{f}_j$, while encoding many-body correlations through the message passing.
	
	Second, the many-body dressed features are mapped into $N_{\det}$ sets of nucleonic and mesonic orbitals,
	\begin{equation}\label{eq:orbitals}
		\begin{aligned}
			\phi_{i\mu}^{(n)} &= O_N^{(n)}(\boldsymbol{f}_i)_\mu\,\exp\!\bigl(-{\xi^{(n)}}^2\,\bar r_i^{\,2}\bigr),\\
			\varphi_M^{(n)} &= O_M^{(n)}(\boldsymbol{f}_M)\,\exp\!\Bigl(-\sum_{\alpha=1}^{3}{\eta_\alpha^{(n)}}^2\,\bar r_{M,\alpha}^{\,2}\Bigr),
		\end{aligned}
	\end{equation}
	where $O_N^{(n)}$ and $O_M^{(n)}$ are residual networks, and the Gaussian envelopes ensure normalizability of the wave function. The learnable decay rates $\xi$ and $\eta$ adapt to the typical size of each system during training, with the mesonic envelope kept anisotropic to accommodate non-spherical tails of the meson distribution~\cite{Zhang:2025okd}.

	Combining all components, the trial wave function reads
	\begin{equation}\label{eq:Psi_total}
	  \Psi = e^{J}\sum_{n=1}^{N_{\det}} w_n\,\det\!\bigl[\phi^{(n)}\bigr]\,\varphi_M^{(n)},
	\end{equation}
	where $w_n$ are learnable mixing coefficients. Multiple determinants are found to be essential for accurately describing larger nuclei with non-trivial shell structure~\cite{Gnech:2021wfn}. Moreover, the $z$ components of the total spin $S$ and isospin $T$ are fixed as $S_z=S$ and $T_z=T$.  All neural networks in the architecture use $\tanh$ as the activation function. The pure-nucleon ansatz used for the bare nuclei is recovered by removing the meson inputs $\boldsymbol{x}_M, \boldsymbol{x}_{iM}$ and the meson orbital $\varphi_M^{(n)}$.

	The network parameters are optimized variationally by minimizing the loss function $\mathcal{L}=\langle H\rangle$, evaluated with Monte Carlo samples drawn from $|\Psi|^2$ via the Metropolis--Hastings algorithm. The parameters are then updated using the stochastic reconfiguration method~\cite{Sorella:2005zz}. Further details on the training procedure can be found in Ref.~\cite{Wu:2025wvv}. The codes are developed on top of the NetKet package~\cite{Carleo2019,Vicentini2022}.

	\section{Results and discussion}
	We first benchmark the NN-VMC framework on the nuclei $^2$H, $^3$H, $^3$He, $^4$He, $^6$Li, $^8$Be, and $^{12}$C. The ground-state energies, computed with both the lattice nucleon mass $m_N=954$~MeV and the physical value $m_N=938.9$~MeV, together with total-spin expectation values are collected in Table~\ref{tab:nuclear_energies}. The statistical errors of the energies are below 0.1$\%$, demonstrating the high numerical precision of the framework. The wave functions reproduce the exact spin quantum numbers $\langle S^2\rangle = S(S+1)$ for $^2$H, $^6$Li ($S=1$), $^3$H, $^3$He ($S=1/2$), and $^4$He, $^8$Be, $^{12}$C ($S=0$). With the physical nucleon mass, the energies for $A\leq 4$ are in close agreement with the experiment, while $^6$Li and $^{12}$C show a slight under-binding. This under-binding is a known feature of the LO $\slashed{\pi}$EFT, which produces excessive repulsion in $p$-shell systems~\cite{Gnech:2023prs}.  Our calculation gives $E(^8\mathrm{Be})>2E(^4\mathrm{He})$, which is in agreement with the unbound nature of $^8$Be. Since the two-cluster scattering state cannot be well described by our trial wave function designed for bound states, the ground-state energy stalls slightly above the threshold. The lattice results are systematically lower than the physical ones, because the heavier nucleon mass reduces the kinetic energy and deepens the binding.
    \begin{table}[htbp]
    	\centering
    	\caption{Nuclear ground-state energies and $S^2$ expectation values.
    		$E$ uses the lattice nucleon mass $m_N=954$ MeV, while $E_{\text{phys}}$ uses the physical $m_N=938.9$ MeV.
    		Statistical errors are shown in parentheses.
    		Experimental values from~\cite{Wang:2021xhn} are listed for comparison.}
    	\label{tab:nuclear_energies}
    	\begin{tabular*}{\hsize}{@{}@{\extracolsep{\fill}}lcccc@{}}
    		\hline\hline
    		Nucleus & $E$ (MeV) & $E_{\text{phys}}$ (MeV) & $\langle S^2\rangle$ & Exp.\ (MeV) \\
    		\hline
    		$^2$H    & $-2.416(0)$    & $-2.240(1)$    & $2.00$ & $-2.224$ \\
    		$^3$H    & $-8.921(1)$    & $-8.470(3)$    & $0.75$ & $-8.482$ \\
    		$^3$He   & $-8.250(1)$    & $-7.808(2)$    & $0.75$ & $-7.718$ \\
    		$^4$He   & $-29.072(1)$   & $-28.161(3)$   & $0.00$ & $-28.30$ \\
    		$^6$Li   & $-32.478(4)$   & $-31.134(12)$  & $2.01$ & $-31.99$ \\
    		$^8$Be   & $-57.873(7)$   & $-55.689(26)$  & $0.02$ & ---      \\
    		$^{12}$C & $-91.621(16)$  & $-87.724(28)$  & $0.03$ & $-92.16$ \\
    		\hline\hline
    	\end{tabular*}
    \end{table}

	\begin{figure}[htbp]
		\centering
		\includegraphics[width=0.9\columnwidth]{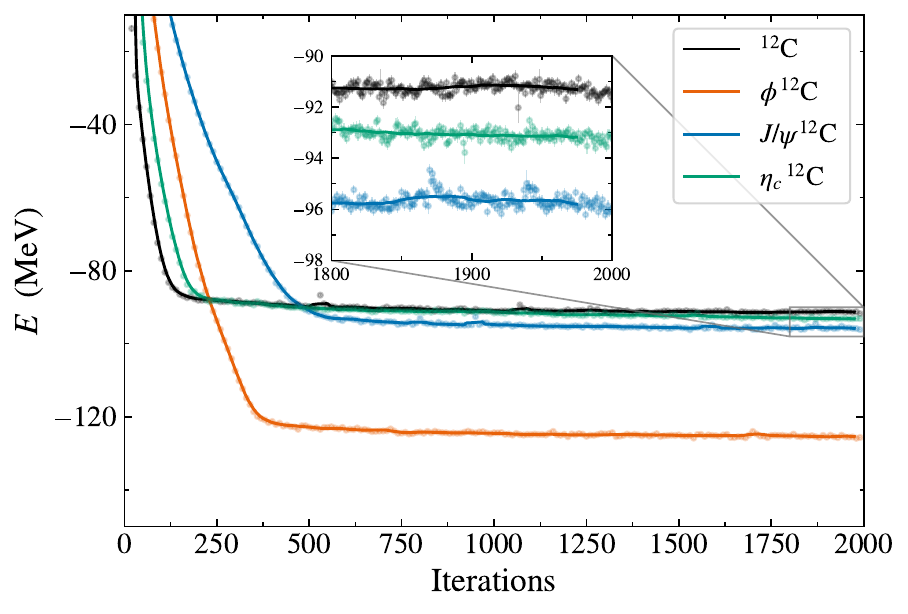}
		\caption{Energy convergence for meson-$^{12}$C system.
			Scatter points with error bars represent the raw VMC energy estimates
			and their Monte Carlo statistical uncertainties; solid curves are
			50-step moving averages, which suppress stochastic fluctuations and highlight the convergence trend.}
		
		\label{fig:convergence}
	\end{figure}
    We now turn to the meson-nucleus bound states. Energy convergence of the training process is shown in Fig.~\ref{fig:convergence}, taking $M$-$^{12}$C ($M=\phi,J/\psi,\eta_c$) as an example.  The ground-state binding energies $E_B=E_{M\text{-}A}-E_A$, total-spin expectation values and root-mean-square (rms) radii are collected in Table~\ref{tab:meson_results}. For each bound state two uncertainties are quoted: the parenthetical $\sigma_{\rm stat}$ is the VMC statistical error, while $\pm\sigma_{V}$ denotes the uncertainty propagated from the HAL QCD potential parameters (see Appendix~\ref{appsec:uncertainty} for details). 
    
    None of the three mesons forms a bound state with a single nucleon. As $A$ increases, the meson-nucleus attraction grows stronger and bound states emerge. The $\phi$-$A$ systems are already bound at $A=2$, with $E_B$ deepening from $1.4$~MeV at $A=2$ to $34$~MeV at $A=12$. The $J/\psi$-$A$ systems are shallowly bound for $A\geq 4$. The existence of the $\eta_c$-$^4$He bound state remains inconclusive given the potential uncertainty; $\eta_c$-$A$ binding is firmly established only for $A\geq 6$, and even $\eta_c$-$^{12}$C is bound by merely $\sim 2$~MeV. The hierarchy $|E_B^{\phi}|\gg |E_B^{J/\psi}|>|E_B^{\eta_c}|$ reflects the relative strengths of the meson-nucleon interactions, consistent with the ordering of the scattering lengths $|a_0^{\phi N}|\gg |a_0^{J/\psi N}|>|a_0^{\eta_c N}|$~\cite{Lyu:2022imf,Lyu:2024ttm}. It is noteworthy that, although bare $^8$Be is an unbound resonance above the $\alpha+\alpha$ threshold, the meson-nucleon attraction is strong enough to stabilize all three $M$-$^8$Be systems below $2E(^4\mathrm{He})$.

\begin{figure*}[htbp]
      \centering
      \includegraphics[width=\textwidth]{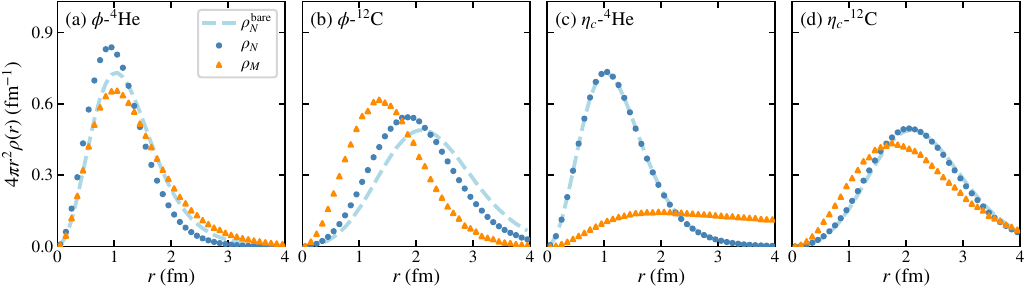}
      \caption{Radial densities for (a) $\phi$-$^4$He, (b) $\phi$-$^{12}$C, (c) $\eta_c$-$^4$He, and (d) $\eta_c$-$^{12}$C. Each panel shows the nucleon radial density (blue circles), meson radial density (orange triangles) of the meson-nucleus system, and the nucleon density of the corresponding parent nucleus (blue dashed line).}
      \label{fig:density_4He_12C}
    \end{figure*}

    \clabel[density]{The spatial configurations of the meson-nucleus systems are characterized by the rms radii in Table~\ref{tab:meson_results} and illustrated by the representative $\phi$- and $\eta_c$-nucleus density distributions in Fig.~\ref{fig:density_4He_12C}.} The nucleon and meson radial densities are defined as
	\begin{equation}
		\begin{aligned}
			& \rho_N(r)=\frac{1}{4\pi r^2}\, \frac{\langle\Psi|\sum_i \delta(|{\boldsymbol{r}}_i-\boldsymbol{R}_{\rm n}|-r)|\Psi\rangle}{A\,\langle\Psi|\Psi\rangle}, \\
			& \rho_{M}(r)=\frac{1}{4\pi r^2}\, \frac{\langle\Psi|\delta(|{\boldsymbol{r}}_M-\boldsymbol{R}_{\rm n}|-r)|\Psi\rangle}{\langle\Psi|\Psi\rangle},
		\end{aligned}
	\end{equation}
    where ${\boldsymbol{R}}_{\rm n}$ is the nuclear c.m. coordinate.  The meson-nucleon rms radii follow the ordering $r_{\phi N}<r_{J/\psi N}<r_{\eta_c N}$ in all systems. In the $M$-$^4$He systems, the $\phi$ is embedded inside the nucleus with $r_{\phi N}\lesssim r_{NN}$, and its density peak coincides with the nucleon density peak; the $\eta_c$ instead forms a halo configuration in which the meson is broadly distributed with a long tail extending well beyond the nucleon distribution. \change{As $A$ increases, the mesons move closer to the nuclear interior, as can be seen by comparing Fig.~\ref{fig:density_4He_12C}(a)(c) with Fig.~\ref{fig:density_4He_12C}(b)(d). The effect is most striking for the $\eta_c$, which evolves from a halo configuration at $A=4$ to a configuration contracted inside the nucleus at $A=12$.} The meson also acts as a ``glue'' analogous to the $\Lambda$ hyperon in light hypernuclei, pulling the nucleons closer to each other~\cite{DiDonna:2025oqf,Zhou:2025anp}: in every meson-nucleus bound state $r_{NN}$ is smaller than $r_{NN}^{\rm bare}$ of the parent nucleus. This contraction is most pronounced in the deeply bound $\phi$-$A$ systems: even for the tightly bound $^4$He, the $\phi$ meson visibly compresses the nucleon distribution, as shown in Fig.~\ref{fig:density_4He_12C}(a), demonstrating that treating $\alpha$ as an inert cluster would miss sizable polarization effects. In the weakly bound states, by contrast, the nucleon distribution remains close to that of the parent nucleus, as seen in Figs.~\ref{fig:density_4He_12C}(c) and (d). \change{The density distributions of other meson-nucleus systems are provided in Appendix~\ref{appsec:density}.}

    \begin{figure}[htbp]
      \centering
      \includegraphics[width=0.8\columnwidth]{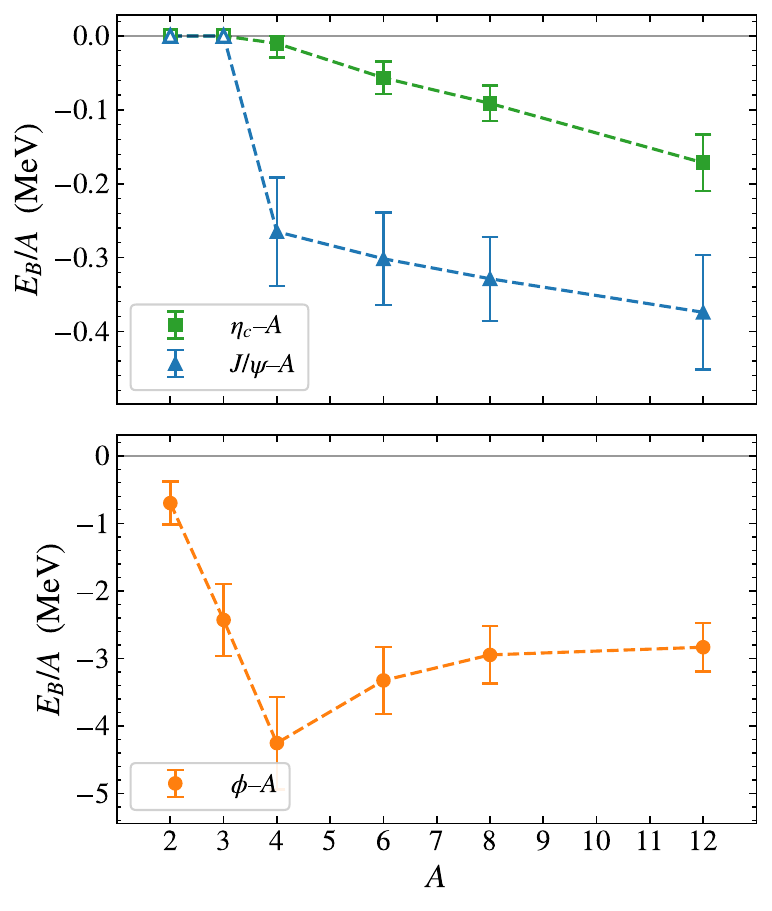}
      \caption{Meson-nucleus binding energy per nucleon $E_B/A$ as a function of nuclear mass number $A$ for the $\eta_c$-$A$, $J/\psi$-$A$ (top) and $\phi$-$A$ (bottom) systems.}
      \label{fig:binding_per_A}
    \end{figure}

    Figure~\ref{fig:binding_per_A} shows the meson-nucleus binding energy per nucleon $E_B/A$, which measures the average meson-nucleon interaction strength. For the charmonium-nucleus systems, $E_B/A$ deepens nearly linearly with $A$ from $A=4$ to $A=12$. By contrast, the $\phi$-$A$ systems exhibit an interesting behavior: $E_B/A$ deepens linearly for $A\leq4$, and then decreases in magnitude toward a saturated value of $\sim-2.8$~MeV.
    These trends can be understood from the spatial structure discussed above. Since all three meson-nucleon potentials are dominated by the short-range interaction at $r\lesssim0.5$~fm, as shown in Fig.~\ref{fig:potentials}, the interaction strength is largely determined by the overlap between the meson and nucleon distributions. In the $J/\psi$-$A$ and $\eta_c$-$A$ systems, the meson evolves from outside the nucleus toward its interior as $A$ increases, so that the overlap and the interaction strength grow accordingly. The same holds for the $\phi$-$A$ systems at $A\leq4$. For $A\geq6$, however, the $\phi$ meson already resides inside the nucleus, while the nuclear volume keeps growing with $A$; this dilutes the nucleon density and weakens the interaction strength. The maximum of $E_B/A$ therefore lies at ${}^4$He, the most compact light nucleus, which offers the largest meson-nucleon overlap. This non-monotonic behavior is a distinctive signature of the short-range, strongly attractive character of the $\phi N$ interaction.

    \begin{table*}[htbp]
		\centering
		\caption{Meson-nucleus ground-state binding energies $E_B$ (in MeV), $S^2$ expectation values and rms radii (in fm).
		$r_{NN}^{\text{bare}}$ is the rms inter-nucleon distance of the bare parent nucleus.
		For each bound state, $\sigma_{\rm stat}$ (compact parenthetical on last digit) is the VMC statistical error
		and $\pm\sigma_{V}$ is the potential-parameter uncertainty.
		$E_B$ for $^8$Be systems are referenced to $2E(^4\mathrm{He})$
		  since bare $^8$Be lies above the $\alpha{+}\alpha$ threshold.}
		\label{tab:meson_results}
		\begin{tabular*}{\hsize}{@{}@{\extracolsep{\fill}}lccccccccccccc@{}}
			\hline\hline
			& \multicolumn{4}{c}{$\phi$-$A$} & \multicolumn{4}{c}{$J/\psi$-$A$} & \multicolumn{4}{c}{$\eta_c$-$A$} & \\
			\cline{2-5}\cline{6-9}\cline{10-13}
			Nucleus & $E_B$ & $\langle S^2\rangle$ & $r_{MN}$ & $r_{NN}$ & $E_B$ & $\langle S^2\rangle$ & $r_{MN}$ & $r_{NN}$ & $E_B$ & $\langle S^2\rangle$ & $r_{MN}$ & $r_{NN}$ & $r_{NN}^{\text{bare}}$ \\
			\hline
			$^2$H    & $-1.400(5) \pm 0.640$    & $0.01$ & $4.09$ & $2.83$ & $>0$ & & & & $>0$ & & & & $3.71$ \\
			$^3$H    & $-7.342(8) \pm 1.604$    & $0.75$ & $2.42$ & $2.28$ & $>0$ & & & & $>0$ & & & & $2.85$ \\
			$^3$He   & $-7.236(6) \pm 1.609$    & $0.75$ & $2.45$ & $2.30$ & $>0$ & & & & $>0$ & & & & $2.90$ \\
			$^4$He   & $-17.010(8) \pm 2.737$   & $2.00$ & $1.97$ & $1.99$ & $-1.059(3) \pm 0.294$  & $2.00$ & $3.22$ & $2.24$ & $-0.041(2) \pm 0.077$  & $0.00$ & $7.04$ & $2.28$ & $2.29$ \\
			$^6$Li   & $-19.952(13) \pm 2.991$  & $0.01$ & $2.58$ & $2.97$ & $-1.814(5) \pm 0.377$  & $0.02$ & $3.45$ & $3.50$ & $-0.344(5) \pm 0.133$  & $2.01$ & $4.89$ & $3.69$ & $3.81$ \\
			$^8$Be   & $-23.576(11) \pm 3.402$  & $2.02$ & $2.77$ & $3.11$ & $-2.632(9) \pm 0.455$  & $2.01$ & $3.49$ & $3.65$ & $-0.726(7) \pm 0.193$  & $0.02$ & $4.15$ & $3.81$ & $4.06$ \\
			$^{12}$C & $-34.00(5) \pm 4.30$     & $2.03$ & $2.73$ & $3.18$ & $-4.49(3) \pm 0.93$    & $2.07$ & $3.15$ & $3.47$ & $-2.06(3) \pm 0.46$    & $0.03$ & $3.38$ & $3.51$ & $3.58$ \\
			\hline\hline
		\end{tabular*}
    \end{table*}

    \begin{table}[htbp]
		\centering
		\caption{Binding energies and rms radii for $J/\psi$-$^6$Li,
		  $\phi$-$^6$Li, and $\phi$-$^2$H in different spin channels,
		  evaluated with the central values of the meson-nucleon potential parameters.}
		\label{tab:spin_resolved}
		\begin{tabular*}{\hsize}{@{}@{\extracolsep{\fill}}lcccc@{}}
			\hline\hline
			System & $\langle S^2\rangle$ & $E_B$ (MeV) & $r_{NN}$ (fm) & $r_{MN}$ (fm) \\
			\hline
			$J/\psi$-$^6$Li & $0.02$ & $-1.815(6)$   & $3.50$ & $3.45$ \\
			                 & $2.01$ & $-1.816(6)$   & $3.53$ & $3.48$ \\
			                 & $6.00$ & $-1.739(6)$   & $3.57$ & $3.54$ \\
			\hline
			$\phi$-$^6$Li   & $0.01$ & $-19.952(13)$ & $2.97$ & $2.58$ \\
			                 & $2.01$ & $-19.711(12)$ & $3.01$ & $2.59$ \\
			                 & $6.01$ & $-19.169(12)$ & $3.08$ & $2.65$ \\
			\hline
			$\phi$-$^2$H    & $0.01$ & $-1.400(5)$   & $2.83$ & $4.09$ \\
			                 & $2.00$ & $-0.807(4)$   & $3.00$ & $4.76$ \\
			                 & $6.00$ & $-0.062(2)$   & $3.31$ & $6.69$ \\
			\hline\hline
		\end{tabular*}
    \end{table}

    Finally, we examine the spin splittings of the meson-nucleus bound states, specifically the $\phi$-$^2$H, $\phi$-$^6$Li, and $J/\psi$-$^6$Li systems. In systems with near-degenerate spin states, the optimized neural-network wave function may not be an eigenstate of the total-spin operator $S^2$, a phenomenon known as spin contamination~\cite{andrews1991spin}. To overcome this, we augment the loss function with a spin penalty term,
    \begin{equation}\label{eq:S2_penalty}
      \mathcal{L}=\langle H\rangle \;\rightarrow\; \langle H\rangle + \omega\,\langle S^2\rangle,\quad \omega = 1~\mathrm{MeV}.
    \end{equation}
    The penalty term effectively enlarges the spin splittings, thereby suppressing the contributions from higher-spin states $S>S_z$ and ensuring that the wave function converges to the $S=S_z$ sector. The resulting binding energies and radii in the different spin channels are collected in Table~\ref{tab:spin_resolved}.
	
	In the $J/\psi$-$^6$Li system, the $S=0$ and $S=1$ states are degenerate within statistical precision, and the $S=2$ state lies only $\sim77$~keV above. The near-degeneracy follows from the geometry: the typical range of spin-dependent $V^s_{J/\psi N}$ is less than $0.5$ fm, whereas $r_{MN}\approx3.5$~fm places the meson well outside its range, strongly suppressing $\langle V^s_{J/\psi N}\rangle$. On the other hand, $V^s_{\phi N}$ in Eq.~\eqref{eq:phiN_doublet} is enhanced by the mass ratio $m_{J/\psi}/m_\phi\approx2.95$, and the smaller $r_{MN}\approx2.6$~fm weakens this geometric suppression. As a result, the $\phi$-$^6$Li system shows a larger splitting of several hundred keV. The spin splitting of $\phi$-$^2$H states is the largest among the three systems, and the wave-function configurations in different channels are substantially altered, as reflected by the large variation of $r_{MN}$ in Table~\ref{tab:spin_resolved}.

    \section{Conclusions and outlooks}
    We carry out the first systematic calculations of $\phi$-, $\eta_c$-, and $J/\psi$-nucleus ground states up to mass number $A{=}12$,
    using HAL QCD meson-nucleon potentials at near-physical pion mass~\cite{Lyu:2022imf,Lyu:2024ttm}
    and a LO $\slashed{\pi}$EFT description of the nuclear interaction~\cite{Schiavilla:2021dun}. The $\phi N(^2S_{1/2})$ interaction unavailable in lattice results is obtained by a model of spin-dependent interactions. 
    The many-body Schr\"odinger equation is solved using the NN-VMC framework, with a Slater-Jastrow neural-network quantum state generalized to incorporate mesonic degrees of freedom.
    The framework is benchmarked on light nuclei from $^2$H to $^{12}$C, yielding ground-state energies consistent with experiment.
    In meson-nucleus systems, we observe the hierarchy of mesonic binding energies $|E_B^\phi|\gg|E_B^{J/\psi}|>|E_B^{\eta_c}|$, and obtain the critical mass number for bound state formation:
    $\phi$-nucleus are bound for $A\ge2$, $J/\psi$ for $A\ge4$, and $\eta_c$ for $A\ge6$, with $\eta_c$-$^4$He inconclusive under potential uncertainties. Interestingly, $E_B/A$ deepens nearly linearly with $A$ for the charmonium systems, whereas the $\phi$-nucleus system exhibits a non-monotonic behavior peaking at $^4$He---a hallmark of the short-range and strongly attractive $\phi N$ interaction.  We also compute the rms radii and density distributions to investigate the spatial structures. With increasing $A$, the meson evolves from a halo to a configuration embedded inside the nucleus. The meson also acts as a ``glue'', compressing the nucleon distribution.  Moreover, the spin splittings of $J/\psi$-$^6$Li, $\phi$-$^6$Li, and $\phi$-$^2$H systems are found to be relatively small, which is another characteristic of the short-range meson-nucleon interaction.
    Since the spin-dependent $\phi N$ interaction has not yet been determined in lattice QCD and is obtained from model assumptions, experimental measurements of $\phi$-nucleus spin splittings would directly constrain this interaction.

    Our calculations are performed at near-physical point,  and the uncertainty from extrapolating to physical values is expected to be small. Future experiments at J-PARC~\cite{Ohnishi:2019cif}, Jefferson Lab~\cite{JLab:PAC42}, and FAIR~\cite{Wiedner:2011mf} may provide the first probes of meson-nucleus bound states~\cite{Krein:2017usp,Metag:2017yuh}. A quantitative comparison between theoretical predictions and experimental data will shed light on the underlying meson-nucleon dynamics.

    \section*{Acknowledgments}We are grateful to Yan Lyu for providing HAL QCD potential data and the helpful discussions on the uncertainty analysis. We also thank Lu Meng for the helpful discussions. This project was supported by the National
    Natural Science Foundation of China (Grants No. 12475137 and No.125B2105). The computational resources are supported by High-Performance Computing Platform of Peking University.

    \section*{Data Availability}
    The data will be made available upon request.

\appendix
\section{Uncertainty of binding energies}
\label{appsec:uncertainty}

In this work, two types of uncertainty for the meson--nucleus binding energies are reported:
the VMC statistical error $\sigma_{\rm stat}$,
and the uncertainty propagated from the HAL QCD meson--nucleon potential parameters $\sigma_V$.

In the VMC method, the energy expectation value $\langle E\rangle$ is estimated as the sample mean
$\bar{E} = \frac{1}{N}\sum_{t=1}^N E_{L,t}$ over $N$ Monte Carlo configurations,
where $E_{L,t} = \Psi^{-1} H \Psi$ is the local energy evaluated on configuration $t$.
The statistical error measures the numerical error due to finite and random Monte Carlo sampling. It is estimated using the blocking method. The $N$ samples are divided into $m$ batches of length $l$, the batch-average energies are
\begin{equation}\label{eq:batch_avg}
	\bar{E}^i = \frac{1}{l} \sum_{t=(i-1)l+1}^{il} E_{L,t}, \quad i = 1, \dots, m.
\end{equation}
The statistical error of $\bar{E}$ is defined as
\begin{equation}
	\sigma_{\rm stat}=\frac{\sqrt{\operatorname{var}(\bar{E})}}{\sqrt{m}}=\sqrt{\frac{1}{m^2} \sum_{i=1}^{m} (\bar{E}^i - \bar{E})^2},
\end{equation}
where $\operatorname{var}(\bar{E})$ is the variance of batch means. The batch size $l$ is chosen sufficiently large that the batch means are effectively uncorrelated.
In this work, all statistical errors are below $0.1\%$ of the corresponding energy values,
confirming that the Monte Carlo sample sizes are adequate for precise estimates.

The HAL QCD meson--nucleon potentials~\cite{Lyu:2022imf,Lyu:2024ttm}
are obtained by fitting the parametrized forms
(Eqs.~\eqref{eq:MN_gauss}--\eqref{eq:phiN_quartet}) to lattice QCD data.
The fitting uncertainties are estimated via the jackknife method:
$N_{\rm jk}$ sets of reduced ensemble are generated by removing subsets of gauge configurations, and the fit is repeated to yield $N_{\rm jk}$ sets of potential parameters $\{p^{(k)}\}_{k=1}^{N_{\rm jk}}$. The central parameters and the uncertainties are given by
\begin{equation}
	\begin{aligned}
	&\bar{p}=\frac{1}{N_{\rm jk}}\sum_{k=1}^{N_{\rm jk}}p^{(k)},\\
	&\sigma_{p}=\sqrt{\frac{N_{\rm jk} - 1}{N_{\rm jk}} \sum_{k=1}^{N_{\rm jk}} (p^{(k)} - \bar{p})^2}.
	\end{aligned}
\end{equation}
The neural-network quantum state is optimized to the ground state of $H(\bar{p})$. To estimate the uncertainties propagated to the meson--nucleus binding energies,
we evaluated the energy on each jackknife parameter set $p^{(k)}$ using the optimized neural-network wave function $\Psi$:
\begin{equation}\label{eq:jk_energy}
  E^{(k)} = \frac{\langle\Psi|H(p^{(k)})|\Psi\rangle}{\langle\Psi|\Psi\rangle}, \quad k = 1, \dots, N_{\rm jk}.
\end{equation}
The uncertainty is then given by the jackknife error estimate,
\begin{equation}\label{eq:jk_sigma}
  \sigma_V = \sqrt{\frac{N_{\rm jk} - 1}{N_{\rm jk}} \sum_{k=1}^{N_{\rm jk}} (E^{(k)} - \bar{E}_{\rm jk})^2},\qquad
  \bar{E}_{\rm jk} = \frac{1}{N_{\rm jk}} \sum_{k=1}^{N_{\rm jk}} E^{(k)}.
\end{equation}
This approach assumes that the wave function does not change appreciably under the potential variations, which is justified by the small uncertainty of the potentials. 
We have also verified this assumption by performing full re-optimizations for selected jackknife parameter sets,
finding consistent results with the estimates. The jackknife method automatically preserves the correlation among the fit parameters,
whereas a naive analysis assuming independent parameter distribution would overestimate the uncertainty by a large factor. The $\phi N({}^2S_{1/2})$ potential in Eq.~\eqref{eq:phiN_doublet} depends on parameters of both $\phi N({}^4S_{3/2})$ and $J/\psi N$ potential.
For simplicity, the jackknife uncertainty is evaluated by varying only the $\phi N({}^4S_{3/2})$ parameters,
while keeping the $J/\psi N$ parameters at their central values. The $N_{\rm jk}$ sets of parameters are provided by the authors of Ref.~\cite{Lyu:2022imf,Lyu:2024ttm}.

In addition to the uncertainty, we also briefly discuss the variance of the local energy, 
\begin{equation}
\sigma^2 = \frac{\langle \Psi |(H-\langle E\rangle)^2|\Psi\rangle}{\langle \Psi|\Psi\rangle}=\frac{1}{N}\sum_{t=1}^N (E_{L,t}-\bar{E})^2.
\end{equation}
The variance measures the fluctuation of $E_L$ over the Monte Carlo samples. For an exact eigenstate, $E_L$ is constant in the configuration space and $\sigma^2 = 0$ ; a small $\sigma^2$ therefore indicates that the wave function closely approximates a true eigenstate.

In the present calculations, we find that the short-range, strongly attractive meson--nucleon potentials---particularly the $\phi N$ interaction---give rise to a large variance.
This is a consequence of the sampling method: the Monte Carlo samples are distributed by $|\Psi|^2$, and the short-range sampling is suppressed by the $r^2$ factor in the radial volume element $d^3r=4\pi r^2 dr$. As a result, the wave function is less thoroughly optimized in the short-range regions~\cite{Keeble:2019bkv}, where the meson-nucleon potential is strongest, and $E_L$ may deviate substantially from $\langle E\rangle$. These few deviant short-range samples result in a large variance.
Nevertheless, the short-range contributions to the expectation values are also suppressed by the $r^2$ geometric factor in the integral. Therefore, the binding energies reported in this work are not appreciably
biased by the inflated variance and remain reliable.

\section{Meson-nucleus density distribution}
\label{appsec:density}
\begin{figure*}[htbp]
  \centering
  \includegraphics[width=0.9\textwidth]{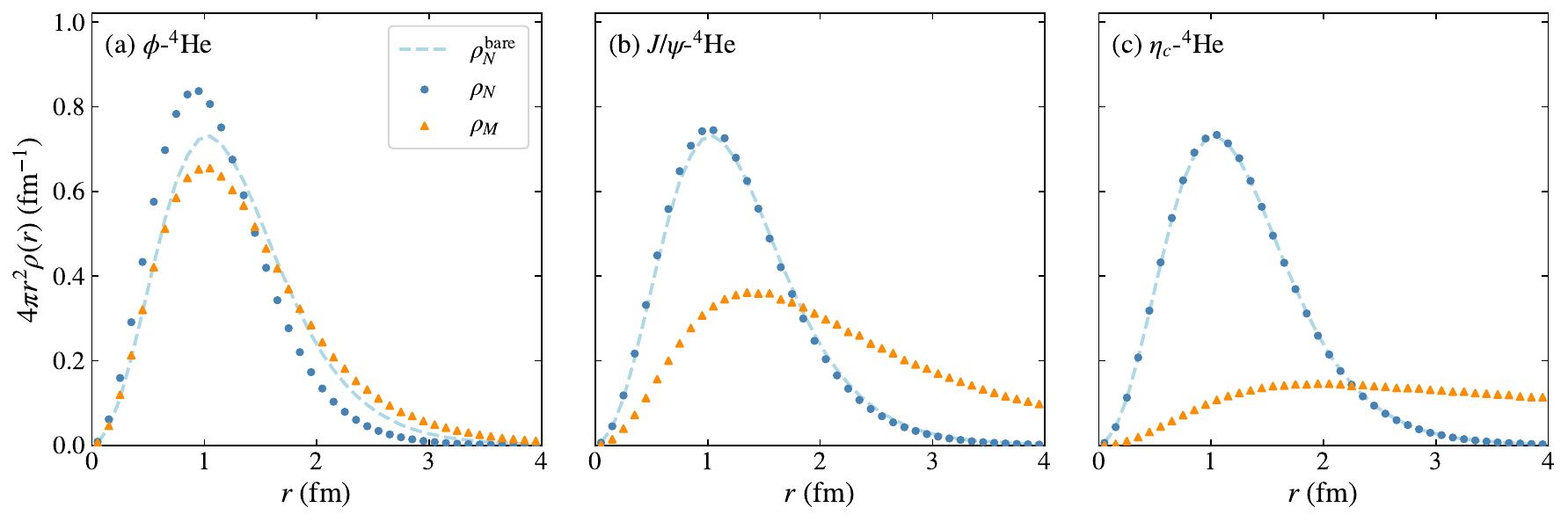}
  \includegraphics[width=0.9\textwidth]{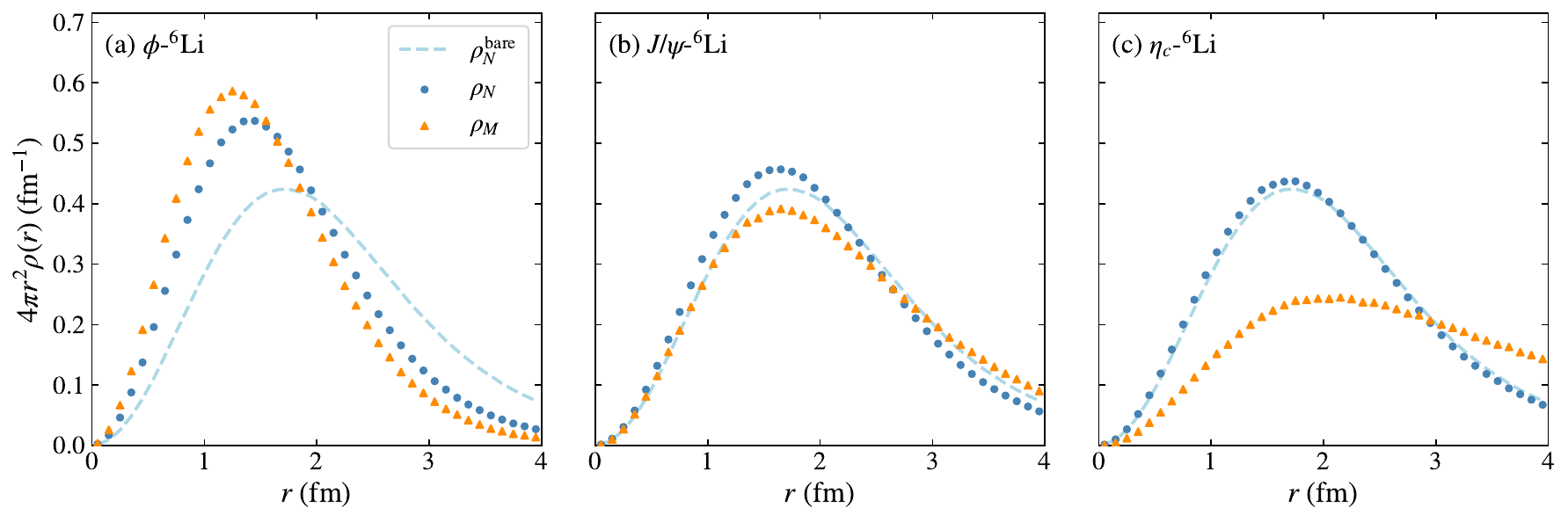}
  \includegraphics[width=0.9\textwidth]{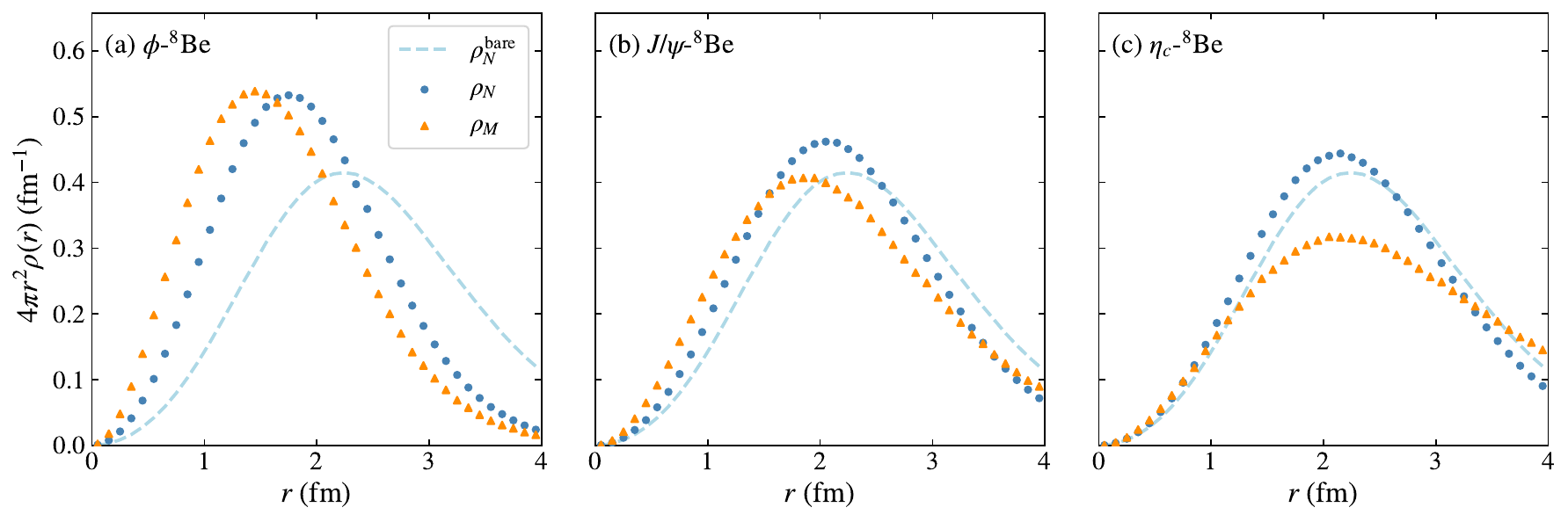}
  \includegraphics[width=0.9\textwidth]{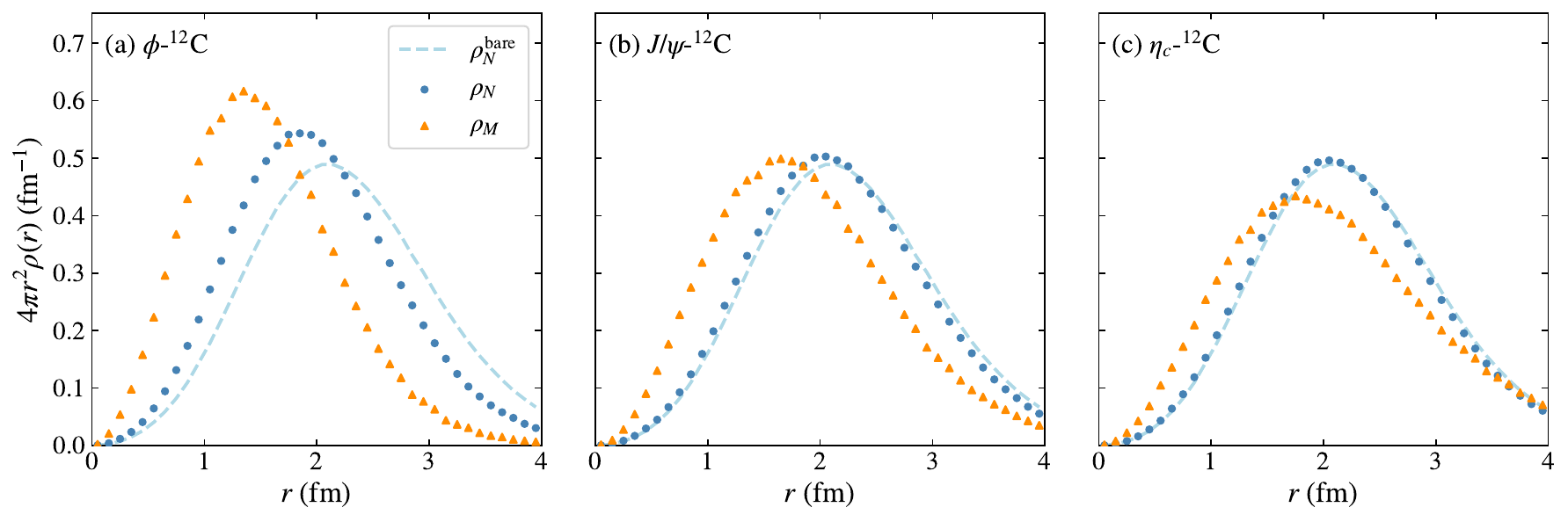}
  \caption{Radial density of the $M$-$^4$He, $M$-$^6$Li, $M$-$^8$Be, and $M$-$^{12}$C. Each panel shows the nucleon radial density (blue circles), meson radial density (orange triangles) of the meson--nucleus system,
  	and the nucleon density in the parent nucleus (blue dashed line).}
  \label{fig:density}
\end{figure*}

Fig.~\ref{fig:density} shows the radial density distributions of the $M$-$^4$He, $M$-$^6$Li, $M$-$^8$Be, and $M$-$^{12}$C systems. With increasing $A$, all three mesons move closer to the nuclear interior.

\bibliography{refs}

\end{document}